\documentclass[10pt,twocolumn]{article}

\usepackage{lipsum} 		
\usepackage{blindtext} 		
\usepackage{amsmath}
\usepackage{amsfonts}
\newcommand{\intprod}{\mathbin{\raisebox{\depth}{\scalebox{1}[-2]{$\lnot$}}}}

\usepackage[superscript, biblabel]{cite}  	

\makeatletter
\renewcommand\@biblabel[1]{#1.}
\makeatother

\usepackage{etoolbox}
\patchcmd{\thebibliography}{\section*{\refname}}{}{}{}

\usepackage[margin=1.0in,hmarginratio=1:1,top=32mm,columnsep=15pt]{geometry}

\usepackage{color}				
\definecolor{dark-gray}{gray}{0.1}	
\usepackage{times}				
\usepackage{microtype} 			
\usepackage[english]{babel} 		

\usepackage[hang, font={footnotesize}, labelfont={bf,up}, textfont={sf,up}]{caption} 
\usepackage{booktabs} 	
\usepackage{mwe}  		

\usepackage{enumitem} 		
\setlist[itemize]{noitemsep} 	

\usepackage{abstract} 
\AtBeginDocument{}  		

\usepackage{titlesec} 			
\renewcommand\thesection{\Roman{section}.} 		  		
\renewcommand\thesubsection{\thesection\Alph{subsection}.} 	
\renewcommand\thesubsubsection{\thesubsection\arabic{subsubsection}.} 
\titleformat{\section}[block]{\normalfont\sffamily\bfseries}{\thesection}{1em}{\MakeUppercase}{} 	
\titleformat{\subsection}[block]{\normalfont\sffamily\bfseries}{\thesubsection}{1em}{}{}  
\titleformat{\subsubsection}[block]{\normalfont\sffamily\bfseries}{\thesubsubsection}{1em}{}{}  
\titlespacing*{\section}{0.0em}{1em}{0.25em}		
\titlespacing*{\subsection}{0.0em}{1em}{0.25em}	
\usepackage{indentfirst}						

\usepackage{fancyhdr} 

\fancypagestyle{plain}{
\fancyhf{}
\lhead{\color{dark-gray}\textit{Submitted to ANS/NT (2021)}} 
\rhead{LA-UR-20-30042 (v3)} 
}

\usepackage{titling} 		

\usepackage{hyperref} 	
\hypersetup{
	backref=true,       
    	pagebackref=true,               
    	hyperindex=true,                
    	colorlinks=true,                
    	breaklinks=true,                
    	urlcolor= black,                
    	linkcolor=blue,                
    	bookmarks=true,                 
    	bookmarksopen=false,
    	filecolor=black,
    	citecolor=blue
}

\pretitle{\begin{center}\large\bfseries} 	
\posttitle{\end{center}} 				
\title{\vspace{-0.3in} \sffamily{On the Symmetry of Blast Waves} }	
\author{%
\normalsize Roy S. Baty (XTD-PRI)\thanks{corresponding author: rbaty@lanl.gov}\, and Scott D. Ramsey (XTD-NTA) \\[-0.5ex] 
\normalsize Applied Physics Theoretical Design Division \\[-0.5ex] 
\normalsize Los Alamos National Laboratory
}
\date{ } 
\renewcommand{\maketitlehookd}{%
\begin{abstract}
\vspace*{-0.5in}
\noindent {\normalfont\textbf{Abstract}\textemdash}
\noindent 
This article presents a brief historical review of G.\,I.\,Taylor's solution of the point blast wave problem which was applied to the Trinity test of the first atomic bomb. Lie group symmetry techniques are used to derive Taylor's famous two-fifths law that relates the position of a blast wave to the time after the explosion and the total energy released. The theory of exterior differential systems is combined with the method of characteristics to demonstrate that the solution of the blast wave problem is directly related to the basic relationships that exist between the geometry (or symmetry) and the physics of wave propagation through the equations of motion. The point blast wave model is cast in terms of two exterior differential systems and both systems are shown to be integrable with local solutions for the velocity, pressure, and density along curves in space and time behind the blast wave.
\begin{center}
\par
\noindent Endorsement:  Len G. Margolin (XCP-5)
\vspace*{-0.05in}
\end{center}
\end{abstract}
}

\begin{document}

\maketitle	


\section{Historical Introduction}

In 1950, the Proceedings of the Royal Society of London published two papers by Sir Geoffrey Taylor [1] and [2]: the main purpose of these papers was to develop, solve, and apply the results of a so-called point blast wave problem to the determination of the energy released in an intense explosion. These papers are of enduring historical significance for many reasons; and not the least of which being, in the second paper, Taylor's provision of an estimate of the yield of the explosion of the Trinity atomic test conducted July 16, 1945. As noted by Barenblatt [3],

\bigskip

``... Taylor's prediction of the value ... caused, in his words, `much embarrassment' in American government circles ..."

\bigskip

\noindent{as the yield of the Trinity explosion was still, at the time, a closely guarded secret; even though the length and time-stamped high-speed photographic stills of the Trinity fireball (owing to Julian Mack, among others) used as an essential component of Taylor's analysis had otherwise been declassified and released to the public, Fig.\,1.}

Taylor compared his solution of the evolution of the blast wave to the published photographs to estimate the initial energy release as a function of the specific heat ratio for an ideal gas, \( \gamma \). The blast wave analysis produced the following yield estimates: 34 kilotons for \( \gamma = 6/5 \), 22.9 kilotons for \( \gamma = 13/10 \), 16.8 kilotons for \( \gamma = 7/5 \), and 9.5 kilotons for \(  \gamma = 5/3 \). The historical yield of the Trinity test published by the U.S.\,Department of Energy is 21 kilotons, [4]. The modern estimate of the yield of the Trinity test is \( 24.8 \  (\pm 1.98) \) kilotons, which is presented in this issue by Selby, et al.\,[5]. Given the physical complexity of the atomic explosion, the fact that Taylor's blast wave solution accurately bounded and predicted the Trinity yield is remarkable. 

\begin{figure*}[ht]
\begin{center}
\includegraphics[height=6cm]{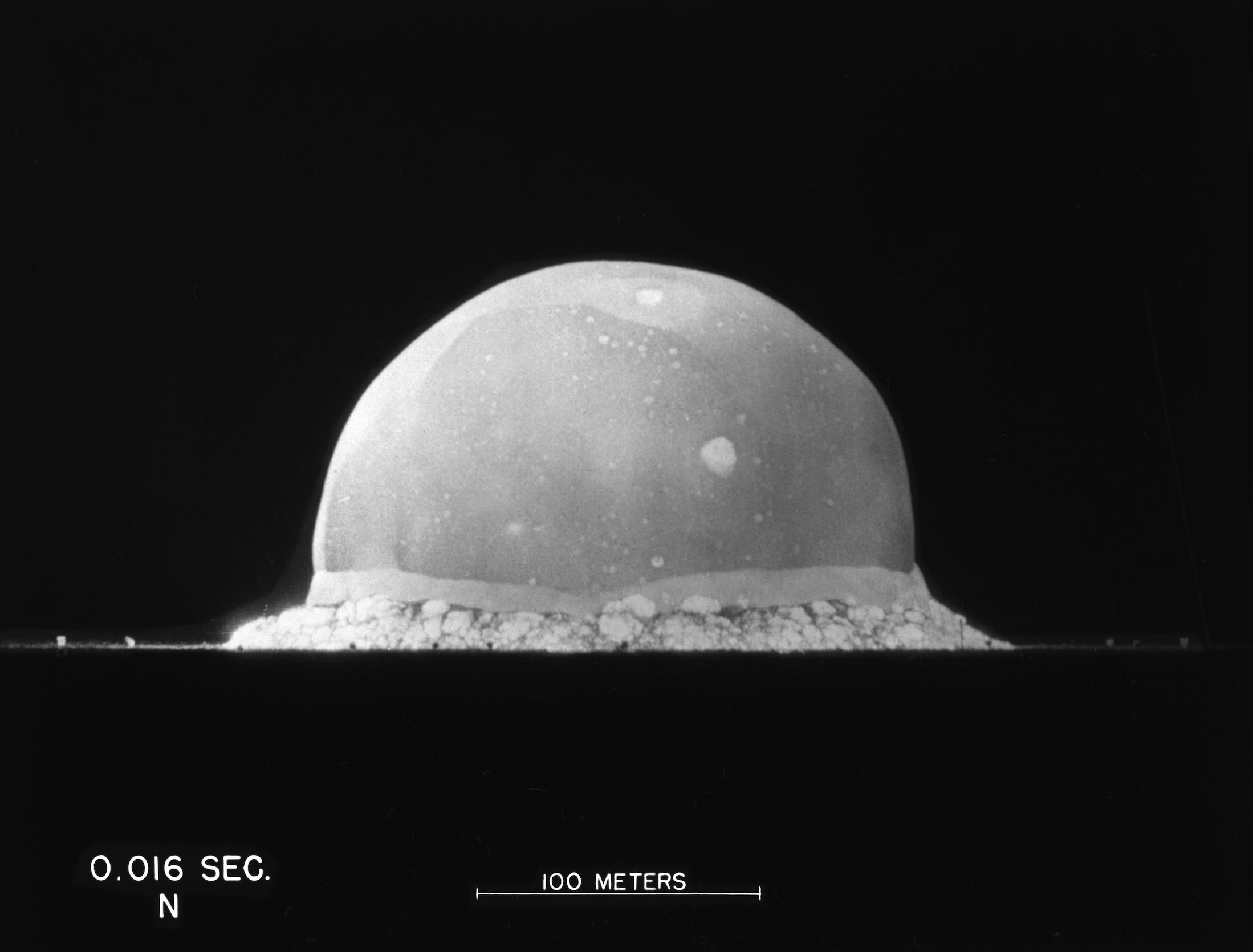} 
\caption{\label{fig:wide} Photograph of the Trinity atomic test showing a blast wave.}
\end{center}
\end{figure*}

The historical details surrounding the genesis of the intense explosion model are fascinating in their own right, though the full story is too lengthy to be reproduced in this short note. A full account of the effort is provided by Clark [6]; see also the books by Barenblatt [3] and [7] for shorter summaries and attendant technical insight into the implications of the work. A discussion of Taylor's technical contributions to the blast wave problem in contrast with parallel analyses is discussed by Deakin [8]. However, by all accounts, a central question surrounding the investigation involved the potential mechanical effects associated with an intense explosion, namely the generation of a powerful blast (or shock) wave. 
Given the well-established difficulties associated with solving any set of fluid mechanical equations involving shock waves, again according to Barenblatt [3],

\bigskip

``... in the whole of Britain there was only one man able to solve this problem -- Professor G.\,I.\,Taylor."

\bigskip

Indeed, given his expertise in fluid mechanics honed over the course of nearly three decades, Taylor recognized that a mathematical model of the minimum required fidelity for characterizing shock wave propagation in air is represented by the spherically symmetric compressible Euler equations in an ideal gas, namely
\begin{equation}
\rho_t + u \rho_x + \rho u_x + 2\frac{\rho u}{x} = 0,
\end{equation}
\begin{equation}
\rho u_t + \rho u u_x + p_x = 0,
\end{equation}
\begin{equation}
p_t + u p_x + \gamma p \Big(u_x + 2\frac{u}{x}\Big) = 0,
\end{equation}
%
%
where \( x \) and \( t \) denote the spherical radial position coordinate and time, respectively, \( \rho \), \( u \), and \( p \) denote the fluid density, radial flow velocity, and isentropic pressure, respectively, subscripts denote partial derivatives with respect to the indicated variables, and \( \gamma \) is the specific heat ratio of an ideal gas, further assumed to be a given constant for air. By the 1940s it was already well established that in the absence of dissipative mechanisms Eqs.\ (1) to (3) admit weak or piecewise continuous solutions, with discontinuities that are interpreted as shock waves of infinitesimal width. 

In shock wave solutions of Eqs.\ (1) to (3), the differential equations themselves are taken to hold on either side of the discontinuity, but not at the location of the discontinuity itself, \( x = R(t) \). To ensure conservation of mass, momentum, and energy across this discontinuity, the ideal gas Rankine-Hugoniot jump conditions must be enforced at the boundary of the blast wave (\( x = R(t) \)), given by 
%
%
\begin{equation}
\rho_0 (u_0 - R_t ) = \rho_1 \Big(u_1 - R_t \Big),
\end{equation}
\begin{equation}
p_0 + \rho_0 ( u_0 - R_t )^2= p_1 + \rho_1 ( u_1 - R_t )^2,
\end{equation}
\begin{multline}
\frac{\gamma p_0}{(\gamma -1)\rho_0} + \frac{1}{2}\Big( u_0 - R_t \Big)^2 \\
= \frac{\gamma p_1}{(\gamma -1)\rho_1} + \frac{1}{2}\Big( u_1 - R_t \Big)^2,
\end{multline}
where the subscripts 0 and 1 denote, respectively, the unperturbed and perturbed fluid state immediately adjacent to the shock wave location, and \( R_t \) is the shock wave speed.

In the 1940s, even Eqs.\ (1) through (6) -- themselves a drastic simplification of vastly more complicated entities such as the multi-dimensional compressible Navier-Stokes equations -- proved impossible to solve in general (either analytically or numerically) without resorting to a variety of additional simplifying assumptions. At this point, as noted by Barenblatt [3],

\bigskip

``... G.\,I.\,Taylor, however, was astute. His ability to deal with seemingly unsolvable problems, by apparently minor adjustment converting them to problems admitting simple and effective mathematics, was remarkable ..."

\bigskip

Indeed, Taylor's astonishing physical insight allowed him to formulate the blast wave problem as depicted in Fig.\,2. In this scenario, for \( t \ge 0 \) the air exterior to the blast wave is assumed to be quiescent and of constant density, namely
\begin{equation}
u_0 = u(x > R) = 0,
\end{equation}
\begin{equation}
\rho_0 = \rho(x > R),
\end{equation}
where \( \rho_0 \) is a fixed positive constant; these conditions present little conceptual difficulty. However, perhaps less intuitively obvious are Taylor's two additional key assumptions, including for \( t \ge 0 \),
\begin{equation}
p_0 = p(x > R) = 0,
\end{equation}
indicating that the ambient counter-pressure \( p_0 \) of the air into which the blast wave propagates is negligible in comparison to the pressure in the region bounded by the blast wave. 

\begin{figure*}[ht]
\begin{center}
\includegraphics[height=6cm]{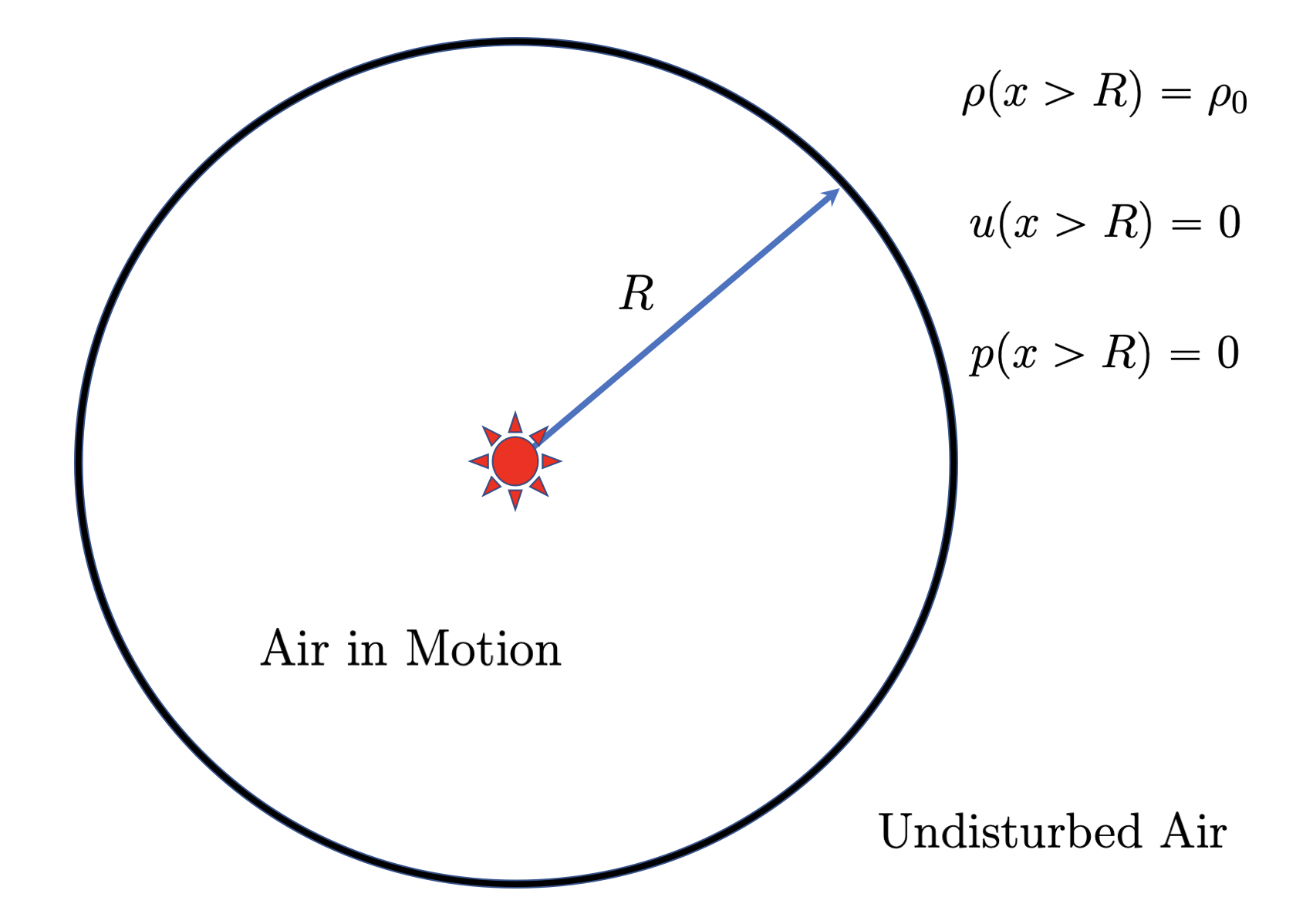} 
\caption{\label{fig:wide} Diagram defining Taylor's point blast wave problem.}
\end{center}
\end{figure*}

Finally, Taylor assumed the initial length scale \( R_0 \) of the rapidly exploding object is negligible in comparison to the spherical blast wave radius \( R \) evaluated at some later time, so that the finite initial blast energy \( E_0 \) (or yield) is assumed to be instantaneously released at an infinitesimal point in space and time (\( x = 0 \) and \( t = 0\)). Among other consequences this assumption results in the formulation of the so-called energy integral given by
\begin{equation}
E_0 = 4\pi \int_{0}^{R} \Big( \frac{p}{\gamma -1}+\frac{1}{2}\rho u^2\Big)x^2 dx,
\end{equation}
indicating the total energy interior to \( R \) is the initial blast energy \( E_0 \), and is conserved for all times. This final assumption also lends itself to the more precise name of the model depicted in Fig.\,2, {\emph{Taylor's point blast wave problem.}}

With Eqs.\ (1) through (9) and the point blast wave assumption, Taylor famously used dimensional analysis techniques to reduce and ultimately produce a numerical solution of Eqs.\ (1) to (6) for not only the blast wave trajectory, \( R(t)\), but also the space time distributions for the velocity, \( u \), pressure, \( p \), and density, \( \rho \), for the flow field at positions behind the blast wave \( x < R(t) \). Consistent with the techniques of dimensional analysis, and among many other remarkable characteristics, Taylor's solution for the flow field behind the blast wave has the properties of being both scale-invariant and self-similar. Figure 3 shows the similarity solutions for velocity, pressure, and density computed by Taylor, [1]. 

It should also be noted that both John von Neumann, [9] and Leonid Sedov, [10] independently produced closed-form solutions to the point blast wave problem, as a direct consequence of their explicit identification and use of Eq.\ (10). As such, the scenario depicted in Fig.\,2 is now typically referred to as the Sedov-Taylor-von Neumann point blast wave (or point explosion) problem; for additional details see Korobeinikov, [11], or Kamm, [12].

\begin{figure*}[ht]
\begin{center}
\includegraphics[height=6cm]{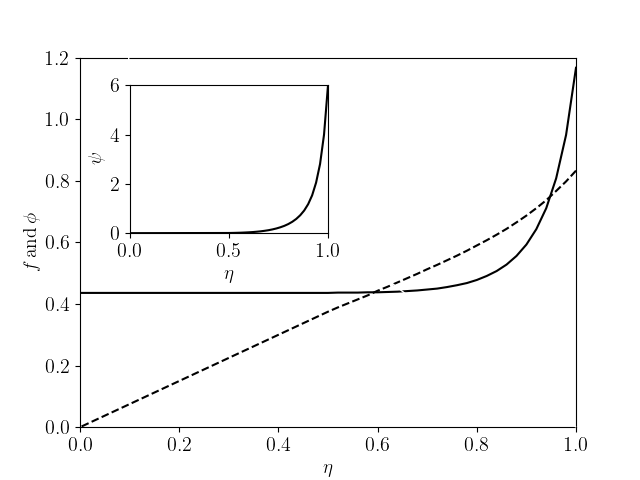} 
\caption{\label{fig:wide} Taylor's solution of the point blast wave in terms of the similarity variable, \( \eta \). In the large plot, the dotted curve is the velocity, \( \phi \), and the solid curve is the pressure, \( f \), while the inserted plot is the density, \( \psi \).}
\end{center}
\end{figure*}

\section{Symmetry Analysis}

With implicit relevance to Taylor's analysis of the blast wave, Garrett Birkhoff published his classic work \emph{Hydrodynamics: A Study in Logic, Fact, and Similitude}, [13], also in 1950. With much reference to an earlier seminal text on dimensional analysis by Percy Bridgman [14], one of the stated objectives of Birkhoff's work is to provide

\bigskip

``... a critical account ... of dimensional analysis. This is usually invoked in justifying model experiments; it has the advantage of requiring no mathematical background beyond high-school algebra, but has the disadvantage of needing additional postulates, whose physical validity must be tested independently. [Birkhoff] give[s] these postulates a group-theoretic formulation, in terms of the dimensional group of all changes of fundamental units."

\bigskip

The ultimate success of Birkhoff's systemization of dimensional analysis and model reduction techniques within the formalized setting of group theory in turn inspired the widespread revitalization of 19th century mathematician Sophus Lie's group-theoretic techniques as a generalized solution strategy for differential equations (see, for example, Ovsyannikov [15], or Olver [16]). The seemingly curious confluence of all these events initially within the context of hydrodynamics is, according to Andreev et al., [17], not to be wondered at: 

\bigskip

``... this analysis was especially fruitful in application to the basic equations of mechanics and physics because the invariance principles are already involved in their derivation. It is in no way a coincidence that the equations of hydrodynamics served as the first object for applying the new ideas and methods of group analysis which were developed by L.V. Ovsyannikov and his school ..."

\bigskip

\noindent{Accordingly, the group-theoretic (or symmetry) interpretation of dimensional analysis and similarity reduction theories has profound consequences for self-similar scaling hydrodynamic phenomena, including the solution of the Taylor point blast wave.}

To this point, a symmetry of a given equation is defined to be a coordinate transformation of the equation such that the equation remains invariant. While the basic goal of symmetry analysis is to determine the coordinate transformations which leave an equation invariant, the practical utility of symmetry analysis is to understand the transformations that may be used to solve or simplify the equation. Let 
\begin{equation}
F(x, u, u_{(1)}, . . .,u_{(k)} ) = 0,
\end{equation}
represent a partial differential equation (PDE) of order \( k \), where \( x = (x_1, . . ., x_n) \) and \( u = (u^1, . . ., u^m)\) are the independent and dependent variables respectively, and \( u_{(l)}\) represents the totality of derivatives of order \( l \). 

Lie studied transformations of the form
\begin{equation*}
(x, u, u_x) \longrightarrow (\tilde{x}, \tilde{u}, \tilde{u}_x),
\end{equation*}
as symmetries, where   
 \begin{equation*}
\tilde{x} = \tilde{x}(x,u,u_x),
\end{equation*}
\begin{equation*}
\tilde{u} = \tilde{u}(x,u,u_x),
\end{equation*}
and
\begin{equation*}
\tilde{u}_x = \tilde{u}_x(x,u,u_x),
\end{equation*}
leave Eq.\,(11) invariant. Lie further assumed that derivatives of \( u \) satisfy the following equation in terms of the total derivative: 
\begin{equation}
du = \sum u_{x_i}dx_i.
\end{equation}
Equation (12) has the modern interpretation as a co-vector or differential one-form
\begin{equation}
\omega = du - \sum u_{x_i}dx_i,
\end{equation}
which may be used to analyze the local geometric properties of Eq.\,(11). From a classical perspective, the expression \( du \) of Eq.\,(12) represents the infinitesimal change of \( u \) under an infinitesimal change of the point on the underlying geometric object where the expression is evaluated. Since geometric infinitesimals are not rigorously defined and because total derivatives may be defined along tangent vectors, \( du \) is interpreted as a differential one-form, Spivak [18]. Differential one-forms are the objects dual to vector fields defined locally on curves and surfaces. The use of differential forms is very powerful in the analysis of systems of PDEs. By admitting higher order exterior forms, smooth systems of PDEs are equivalent to exterior differential systems (EDSs) which may be analyzed with mathematical techniques from differential and algebraic geometry, where an EDS is a set of equations in terms of differential forms, Krasil'shchik and Vinogradov [19], and Vinogradov [20].

The one-dimensional Euler equations, Eqs.\,(1), (2), and (3) may be reduced to an EDS by introducing a transformation which explicitly reduces the PDEs to ordinary differential equations (ODEs), or by introducing a universal space of higher dimension which replaces the derivatives in the PDEs with independent variables and then couples the new variables to the original derivatives through differential forms (total derivatives). Both of these approaches are discussed in this article; the transformation approach to outline solutions of the blast wave problem, and the general approach to show how symmetry is coupled to the physics of blast wave propagation. The notation, concepts, and theorems applied in this article on exterior differential systems follow, Ivey and Landsberg [21].

For Eqs.\,(1), (2), and (3), the following independent variables are introduced:
%
%
\begin{equation*}
x^1 \equiv t, \ \ \ x^2 \equiv x, \ \ \ u^1 \equiv \rho, \ \ \ u^2 \equiv u, \ \ \ u^3 \equiv p,
\end{equation*}
and
\begin{multline*}
u^1_1 \equiv \rho_t, \ \ \ u^1_2 \equiv \rho_x, \ \ \ u^2_1 \equiv u_t, \\ u^2_2 \equiv u_x, \ \ \ u^3_1 \equiv p_t, \ \ \ u^3_2 \equiv p_x.
\end{multline*}
The Euler equations are then cast on the 11 dimensional space \( \mathbb{R}^2 \times \mathbb{R}^3 \times \mathbb{R}^6 \) in terms of \( (x^1, x^2) \times (u^1, u^2, u^3) \times (u^1_1, u^1_2, u^2_1, u^2_2, u^3_1, u^3_2) \) which is called the space of 1-jets, and denoted by \( J^1(\mathbb{R}^2, \mathbb{R}^3)\). Eqs.\,(1), (2), and (3) then become
\begin{equation}
F^1 \equiv u^1_1 + u^2u^1_2 + u^1u^2_2 + 2\frac{u^1u^2}{x^2}, 
\end{equation}
\begin{equation}
F^2 \equiv u^1u^2_1 + u^1u^2u^2_2 + u^3_2,
\end{equation}
and
\begin{equation}
F^3 \equiv u^3_1 + u^2u^3_2 + \gamma u^3(u^2_2 + 2\frac{u^2}{x^2});
\end{equation}
%
%
subject to the differential forms:
\begin{multline}
\sigma^1 = du^1 - u^1_1 dx^1 - u^1_2 dx^2 \ \ \ \text{and} \\ d\sigma^1 = - du^1_1 \wedge dx^1 - du^1_2 \wedge dx^2,
\end{multline}
\begin{multline}
\sigma^2 = du^2 - u^2_1 dx^1 - u^2_2 dx^2 \ \ \  \text{and} \\ d\sigma^2 = - du^2_1 \wedge dx^1 - du^2_2 \wedge dx^2,
\end{multline}
\begin{multline}
\sigma^3 = du^3 - u^3_1 dx^1 - u^3_2 dx^2 \ \ \ \text{and} \\d\sigma^3 = - du^3_1 \wedge dx^1 - du^3_2 \wedge dx^2,
\end{multline}
which are used to define the exterior differential system representing the Euler equations, where \( \wedge \)  is the exterior product for differential forms. The exterior (or wedge) product is a multiplicative operation defined on alternating multilinear mappings of vector spaces (tensors). The mapping \( (\rho, u, p)^t : \mathbb{R}^2 \longrightarrow \mathbb{R}^3\) identifies a smooth surface \( \Sigma \subset J^1(\mathbb{R}^2, \mathbb{R}^3)\) via the jet variables with the coordinate cover \( (x, t,\rho, u, p) \).

Next, consider the space of all differential forms defined on \( \Sigma \), \( \Lambda^{\ast}(\Sigma) \equiv \oplus \, \Lambda^k(\Sigma) \), where \( \Lambda^k(\Sigma) \equiv \Gamma (\Lambda^kT^{\ast}\Sigma) \) is the space of sections of smooth functions defined on the bundle of \( k\)-order differential forms. A subspace \( I \subset \Lambda^{\ast}(\Sigma) \) is an algebraic ideal if it is a direct sum of homogeneous subspaces \( I^k \subset \Lambda^k(\Sigma)\) and it is closed under the wedge product. An algebraic ideal \( I \) is a differential ideal if \( dI \subset I\). An exterior differential system on a surface \( \Sigma \) is a differential ideal \( I \subset \Lambda^{\ast}(\Sigma)\).

The EDS representing the Euler equations is then given by the differential ideal
\begin{multline*}
I = \Big{\{} \alpha \wedge \sigma^1 + \beta \wedge \sigma^2 + \lambda \wedge \sigma^3 + \gamma \wedge d\sigma^1 \\
+ \delta \wedge d\sigma^2  + \phi \wedge d\sigma^3
\vert \ \alpha, \beta, \lambda, \gamma, \delta,\phi \in \Lambda^{\ast}(\Sigma)\Big{\}},
\end{multline*}
defined by the algebraic equations (14), (15), and (16), and the differential forms given by Eqs.\,(17), (18), and (19). Moreover, for the general system of PDEs the space and time variables are assumed to be linearly independent \( dx^1 \wedge dx^2 \ne 0 \).

Infinitesimal symmetries have a natural expression for an EDS. Let \( I \) be an EDS on \( \Sigma \). A vector field \( V \) is an infinitesimal symmetry of \( I \) if \( \mathcal{L}_V \Psi \in I \) for all \( \Psi \in I \). Here \( \mathcal{L}_V \) is the Lie derivative with respect to the vector \( V \).

\section{Symmetry Analysis of the Point Blast Wave Problem}

As discussed in Section 1, Taylor's construction of his point blast wave solution proceeds with the use of dimensional analysis techniques. Alternatively, this solution may also be systematically derived through invariance of Eqs. (1) through (10) under a Lie group of scaling transformations. Both approaches are summarized in the sections to follow. However, ultimately, using group-theoretic techniques Taylor's point blast wave solution may be explicitly categorized according to a symmetry group inherent to Eqs.\,(1) through (10).

\subsection{Dimensional Analysis}

The most important result arising from the codification of dimensional analysis techniques in the late 19th and early 20th centuries is the celebrated ``Buckingham-\(\pi\) Theorem'' (so called due to its explicit statement by Edgar Buckingham in 1914) which, according to Barenblatt [7], states

\bigskip

\noindent{{\bf{Theorem}} \emph{A physical relationship between some dimensional (generally speaking) quantity and several dimensional governing parameters can be rewritten as a relationship between some dimensionless parameter and several dimensionless products of the governing parameters; the number of dimensionless products is equal to the total number of governing parameters minus the number of governing parameters with independent dimensions.}}

\bigskip

In his first paper [1], Taylor uses this powerful result to construct the solution of the point blast wave problem. Referring to Fig.\,2, in this analysis Taylor formulates the blast wave position \( R \) as depending on the initial energy \( E_0 \), the time \( t \) elapsed since the explosion began, and the ambient mass density \( \rho_0\) and ideal gas specific heat ratio \( \gamma \) of the surrounding air; that is
\begin{equation}
R = \phi_0(E_0, t, \rho_0, \gamma),
\end{equation}
where \( \phi_0 \) is a function to be determined of the indicated arguments. The dimensions of the parameters appearing in Eq.\,(20) are given by: \( R \sim \text{L}\), \( E_0 \sim \text{ML}^2 \text{T}^{-2} \), \( t \sim \text{T}\), \( \rho_0 \sim \text{ML}^{-3}\), and \(\gamma \) is dimensionless,
where the set of fundamental mechanical dimensions of L, T, and M refer to units of length, time, and mass, respectively. It is readily verified that the dimensions of \( E_0\), \( t\), and \( \rho_0 \) are linearly independent; that is, the dimensions of any one of these quantities cannot be formed as a linear combination of any of the others. 

Since \( R \) has dimensions of L, the function \( \phi_0 \) must also have dimensions of L; accordingly, Eq.\,(20) may be written as
\begin{equation}
R = S(\gamma ) E^{a_1}_0 t^{a_2} \rho^{a_3}_0,
\end{equation}
where \( a_1, a_2, \) and \( a_3 \) are constants to be determined, and where \( S \) is an arbitrary function of the dimensionless quantity, \(\gamma \). The dimensions of Eq.\,(21) are given by
\begin{equation}
L = (\text{ML}^2\text{T}^{-2})^{a_1}(\text{T})^{a_2}(\text{ML}^{-3})^{a_3}.
\end{equation}
Equation (22) produces three linear algebraic equations for powers of L, T, and M in terms of the exponents \( a_1 \), \( a_2 \), and \( a_3 \), which implies that \( a_1 = 1/5, a_2 = 2/5,\) and \( a_3 = -1/5\). With these values for the exponents, Eq.\,(22) becomes
\begin{equation}
R = S(\gamma ) \Big( \frac{E_0}{\rho_0} \Big)^{\frac{1}{5}}\, t^{\frac{2}{5}},
\end{equation}
or, in terms of the equivalent dimensionless \( \pi \)-group,
\begin{equation}
S(\gamma ) \Big( \frac{E_0}{\rho_0} \Big)^{\frac{1}{5}}\, \frac{t^{\frac{2}{5}}}{R} \ = \ \pi_0 \ \ = \ \ \text{constant}.
\end{equation}
Equation (23) is the famous result derived by Taylor, including the proportionality between the spherical blast wave position \( R \) and the two-fifths power of the time \( t \) elapsed since the explosion began. At this point the function \( S \) remains undetermined. 

A solution of Eqs.\,(1) to (6) subject to Eqs.\,(7) through (9) is required to determine \( S \). Proceeding as with Eq.\,(20), Taylor assumes the radial flow velocity, pressure, and density interior to the spherical blast wave to depend on the parameters:
\begin{equation}
\frac{u}{u_1} = F_{u} (x, R, u_1, \gamma),
\end{equation}
\begin{equation}
\frac{p}{p_1} = F_{p} (x, R, p_1, \gamma),
\end{equation}
\begin{equation}
\frac{\rho}{\rho_1} = F_{\rho} (x, R, \rho_1, \gamma),
\end{equation}
where the necessary dependences on \( E_0 \), \( t \), and \( \rho_0 \) are implicit in those of \( R \), and \( \rho_1 \), \( u_1 \), and \( p_1 \) follow from an algebraic solution of Eqs.\,(4), (5), and (6) subject to Eqs.\,(7) to (9). Equations (7) through (9) suggest that in addition to \( E_0 \), \( t \), and \( \rho_0 \), the flow profiles inside the region bounded by the outgoing blast wave, defined by \( R \), also depend on the Rankine-Hugoniot jump conditions immediately behind the shock wave, as indicated by the subscript 1. 

Following an identical procedure that resulted in Eq.\,(23), Eqs.\,(25)\ through (27) become
\begin{equation}
u = u_1 \phi(\eta, \gamma),
\end{equation}
\begin{equation}
p = p_1f(\eta, \gamma),
\end{equation}
\begin{equation}
\rho = \rho_0 \psi (\eta, \gamma),
\end{equation}
where
\begin{equation}
\eta \equiv \frac{x}{R}.
\end{equation}
The independent variable \( \eta \) is a similarity variable, which may be used with the functions \( \phi, f \), and \( \psi \) to reformulate the PDEs (1) to (3) as three ODEs, while Eqs.\,(4) through (6) combined with Eqs.\,(7), (8), and  (9) may likewise be reformulated as three associated initial conditions. The reduction to a system of ODEs is only possible since the \( x \) and \( t \) dependences featured in the definitions of \( \phi\), \( f \), and \( \psi \) appear only in terms of their combination through \( \eta \). Therefore, a solution of the reduced ODEs for \( \phi \), \( f \), and \( \psi \) is necessarily self-similar: since \( R = R(t), u_1 = u_1(t), p_1 = p(t) \), and \( \rho_1 = \rho_1(t) \) the spatial dependence in all flow variables is structurally invariant and may be obtained between different times by the scaling transformations indicated in Eqs.\,(28) through (31).

Finally, for a fixed value of \( \gamma \), a solution for \( \phi \), \( f \), and \( \psi \) transformed back to that for \( u(x,t)\), \( p(x,t) \), and \( \rho(x,t) \) using Eqs.\,(28) to (31) may be substituted into the energy integral given by Eq.\,(10) together with Eq.\,(23) to yield the previously undetermined function \( S( \gamma ) \) as appearing in Eq.\,(23), and so the point blast wave solution is determined in its entirety. That this solution is achievable under construction of only Eqs.\,(23) and (28) through (31) is at the root of various statements appearing in the literature to the effect of:

\bigskip

``... solutions of the first type possess the property that the ... exponents of \( t \) and \( R \) in all scales are determined either by dimensional considerations or from the conservation laws ..." 

\bigskip

\noindent{as noted by Zel'dovich and Raizer [22]. Indeed, the point blast wave solution is remarkable in the elegance of the purely dimensional arguments that give rise to it. As discussed extensively by Barenblatt [3] and [7], the principal difficulty of the analysis -- and, more remarkable still, Taylor's genius in overcoming it -- is in the selection of the minimum parameter set still of sufficient physical fidelity for establishing the dependencies of \( R, \rho, u \), and \( p\), as represented essentially by \( E_0, t, \rho_0\), and \( \gamma \). In turn, the selection of this parameter set demanded of Taylor a significant amount of judgement so as to credibly ignore possible effects such as variable \( \gamma \) or \( \rho_0\), higher-dimensional effects, and the effects of finite \( p_0 \) or \( R_0 \).} 

For this very reason dimensional analysis techniques have been subject to much misunderstanding and criticism from almost the beginning of their formalization in the late 19th century. Certain pointed questions surrounding the systemization of any such efforts remained largely unanswered even through the middle of the 20th century, when Lie's symmetry analysis techniques were essentially rediscovered, and introduced into much of modern physics.

\subsection{Lie Group Analysis}

Classical treatments of dimensional analysis theory (see, for example, Bridgman [14]) include suggestive terminology such as ``change ratios,'' ``complete equations," and ``dimensional homogeneity." Birkhoff [13] first characterized these and related concepts according to their rigorous, common theme: the notion of invariance or symmetry of a mathematical relationship under a group of scaling transformations. 

In turn, the uses of symmetries of mathematical relationships are based on the geometric interpretation of those same relationships. Put simply, Lie's analysis technique is a systematic means for identifying symmetric coordinate systems under which mathematical relationships assume simpler forms than otherwise originally cast. In this sense, the reduction in complexity ultimately afforded via invariance is best interpreted as being connatural with the symmetry set present in a scenario of interest, as opposed to being a byproduct of (in the case of scaling) nondimensionalization principles. 

As such, and in the context of scaling symmetries, all results of Taylor's point blast wave problem as summarized in Section 3.1 may be derived using Lie's systematic framework, beginning with the fundamental governing equations given by Eqs.\,(1) through (6). Proceeding, in the context of Eqs.\,(1) through (6), the maximal Lie group of all potential scaling transformations in all dimensional variables may be written as 
\begin{multline}
\tilde{x} = e^{\alpha_1 \epsilon}x, \ \ \tilde{t} = e^{\alpha_2 \epsilon}t, \ \ \tilde{\rho} = e^{\alpha_3 \epsilon}\rho, \ \ \tilde{u} = e^{\alpha_4 \epsilon}u, \\
\tilde{p} = e^{\alpha_5 \epsilon}p,
\end{multline}
where the tildes represent transformed variables, \( \epsilon \) is known as the group parameter, and \( \alpha_1, \alpha_2, \alpha_3, \alpha_4\), and \( \alpha_5 \) are constants to be determined. The transformations given by Eqs.\,(32) are referred to a Lie group, as they feature an identity element \( \epsilon = 0 \), the inverse element, and the closure and associativity properties under composition.

The goal of symmetry analysis is to identify the precise instantiation of Eq.\,(32) that leaves a given set of mathematical relationships invariant; that is, the relevant equations assume identical forms in both the original and transformed variable sets. Lie's fundamental achievement in the execution of this technique for differential equations was the realization of this equivalence on a purely local (or infinitesimal) level, using a generalized directional derivative operator \( \mathcal{L}_V \) known as a Lie derivative. In the case of Eq.\,(32), the relevant Lie derivative is defined using the vector field
%
%
\begin{multline}
V = \alpha_1 x \frac{\partial}{\partial x} + \alpha_2 t \frac{\partial }{\partial t} + \alpha_3 \rho \frac{\partial }{\partial \rho} \\
+ \alpha_4 u \frac{\partial }{\partial u} + \alpha_5 p \frac{\partial }{\partial p},
\end{multline}
that may be constructed using a first-order Taylor expansion of Eqs.\,(32) about the identity element. Geometrically, \( V \) is a tangent vector to a surface invariant under the transformations given by Eqs.\,(32), existing in the space where \( (x, t, \rho, u, p) \) are regarded as independent coordinates.
%
%
Indeed, invariance of a mathematical relationship under Eq.\,(33) is then assessed by evaluating
\begin{equation}
\mathcal{L}_V \Psi \in I,
\end{equation}
where each $\Psi \in I$ are, for example, the elements of Eqs.\,(1) through (10) cast as a differential ideal. If, during the course of the analysis of all relevant differential forms, at least one of \( \alpha_1, \alpha_2, \alpha_3, \alpha_4\), or \( \alpha_5\) is revealed to be non-zero, the entire mathematical model under investigation is said to possess a scaling symmetry. Moreover, the infinitesimal form of this symmetry given by Eq.\,(33) may then be used to construct a coordinate system in which the symmetric equations \( \Psi \in I \) assume simpler forms; in the case of scaling symmetry, the accompanying reduction is identical to that resulting from maximal application of the Buckingham-\( \pi \) Theorem.

Proceeding, evaluation of Eq.\,(34) with each \( \Psi \in I \) given by each of Eqs.\,(1) through (3) yields \( \alpha_4 = \alpha_1 - \alpha_2 \) and \( \alpha_5 = \alpha_3 + 2\alpha_4 = \alpha_3 - 2 (\alpha_1 - \alpha_2)\) (see, for example, Ramsey and Baty [23]), so that Eq.\,(33) reduces to
%
%
\begin{multline}
V = \alpha_1 x \frac{\partial}{\partial x} + \alpha_2 t \frac{\partial }{\partial t} + \alpha_3 \rho \frac{\partial }{\partial \rho} \\
+ (\alpha_1 - \alpha_2 )u \frac{\partial }{\partial u} + (\alpha_3 + 2\alpha_1 - 2\alpha_2)p \frac{\partial}{\partial p} ,
\end{multline}
indicating that in the absence of initial, boundary, or otherwise ancillary condition information, Eqs.\,(1) through (3) are dimensionally consistent with the mechanical scalings \( u \sim x/t \) and \( p \sim \rho u^2 \). The presence of three independent free parameters in Eq.\,(35) also reflects the presence of three independent fundamental dimensions (e.g., L, T, and M as appearing in Section 3.1) in the formulation of the fluid mechanical system given by Eqs.\,(1), (2), and (3). 

Furthermore, evaluation of Eq.\,(34) with each \( \Psi \in I \) given by each of Eqs.\,(4) through (6), and the initial condition \( R\vert_{t = 0} = 0 \), yields (see, for example, Giron, et al. [24])
%
%
\begin{equation}
\alpha_2 t R_{tt} + (\alpha_2 - \alpha_1 )R_t = 0 \ \ \ \Rightarrow \ \ \ R = kt^{\frac{\alpha_1}{\alpha_2}},
\end{equation}
for \( k = \) constant, further indicating that scale invariant shock wave propagation proceeds according to power-law behavior in time. The precise power appearing in Eq.\,(36) remains arbitrary as neither \( \alpha_1 \) nor \( \alpha_2 \) is constrained given Eqs.\,(4), (5), and (6), which are dimensionally equivalent to Eqs.\,(1), (2), and (3).

Indeed, at this point, none of Taylor's fundamental assumptions (i.e., Eqs.\,(7) to (10)) have been invoked. To rigorously investigate the consequences of assuming Eqs.\,(7) to (9), evaluating Eq.\,(34) with each \( \Psi \in I \) given by each of these conditions yields
\begin{multline}
(\alpha_1 - \alpha_2)u_0 = 0, \ \ \ \alpha_3 \rho_0 = 0, \\ (\alpha_3 + 2\alpha_1 - 2\alpha_2)p_0 = 0,
\end{multline}
satisfaction of which at least requires \( \alpha_3 = 0 \), since \( \rho_0 > 0\) by the physical problem formulation. Otherwise, \( \alpha_1 \) and \( \alpha_2 \) are not further constrained since \( u_0 = 0 \) and \( p_0 = 0 \) via Eqs.\,(7) and (9), respectively. 

Continuing, evaluating Eq.\,(34) with the final \( \Psi \in I \) given by Eq.\,(10) yields (see, for example, Hutchens [25])
\begin{equation}
k = S\Big( \frac{E_0}{\rho_0} \Big)^{\frac{1}{5}} \ \ \ \text{and} \ \ \ \alpha_1 = \frac{2}{5}\alpha_2,
\end{equation}
so that with Eq.\,(38) and \( \alpha_ 3 = 0 \) resulting from Eq.\,(37), Eq.\,(35) finally becomes
%
%
\begin{equation}
V = \frac{2}{5}\alpha_2 x \frac{\partial}{\partial x} + \alpha_2 t \frac{\partial }{\partial t} 
- \frac{3}{5} \alpha_2 u \frac{\partial }{\partial u} - \frac{4}{5} \alpha_2 p \frac{\partial}{\partial p} ,
\end{equation}
a version of which is also given by Cantwell [26]. Moreover, with Eq.\,(38), Eq.\,(36) becomes Eq.\,(23), which is now demonstrated to arise from the presence of a one-parameter (i.e., \( \alpha_2 \) as written) scaling symmetry of Eqs.\,(1) through (10), in turn given by Eq.\,(39). 

This outcome may be compared with the cases where either \( u_0 \ne 0 \) or \( p_0 > 0 \), or when a finite initial length scale \( R_0 \) appears in the problem formulation. In any of these cases, satisfaction of the appropriate manifestation of Eq.\,(37) (for example) further demands \( \alpha_2 = 0 \), thus entirely nullifying Eq.\,(39) and rendering the problem formulation without scaling symmetry. This result reflects the importance of Taylor's intuition resulting in Eqs.\,(7), (8), (9), and (10). More broadly, it also demonstrates the intuitive phenomenon that one degree of freedom is removed from the three-parameter mechanical scaling group of Eqs.\,(1) to (6) for every fixed dimensional constant (e.g., \( \rho_0 \) and \( E_0\)) also introduced into a specific problem formulation. Consequently, for a self-similar scaling solution of Eqs.\,(1) to (6) to exist, at least one such degree of freedom must be present in Eq.\,(39).

Under this construction, Eq.\,(39) may be used to identify the symmetric coordinates in which Eqs.\,(1) through (10) assume a simpler form and may ultimately be solved. In particular, the similarity variables introduced in Section 3.1 via dimensional considerations are now readily shown to actually be the invariant coordinates of the one-parameter scaling group generated by Eq.\,(39), determined according to
%
%
\begin{multline}
V\mathfrak{H} (x,t,\rho,u, p) = 0 \ \ \ \Rightarrow \\ \frac{dx}{\frac{2}{5}x} = \frac{dt}{t} = \frac{d\rho}{0} = -\frac{du}{\frac{3}{5}u} = -\frac{dp}{\frac{4}{5}p},
\end{multline}
where \( \mathfrak{H} \) is an otherwise arbitrary invariant function of the indicated arguments. The solution of the characteristic equations appearing in Eq.\,(40) features four constants of integration; these constants are identical to the functions \( \phi, f, \psi \), and \( \eta \) defined by Eqs.\,(28) through (31). Following the systematic derivation of these similarity variables, the symmetry reduction associated with Eq.\,(39) is revealed to collapse to ODEs otherwise encountered in Section 3.1; the complete solution of the self-similar Taylor point blast wave problem then proceeds in the same manner as previously disseminated. Through this analysis, the entire Taylor blast wave solution is thus revealed to be a direct manifestation of a one-parameter group of scaling transformations admitted by the problem formulation, namely Eqs.\,(1) through (10). 

\section{Method of Characteristics}

An important theme of using symmetry to analyze a system of PDEs is that of simplifying the system of equations. For the Taylor point blast wave problem, symmetry is applied to reduce the system of PDEs to a system of ODEs, which then allows direct numerical integration of the problem. In compressible fluid mechanics, the symmetries that reduce the PDEs to a system of ODEs are related to the fundamental physics of wave propagation. For unsteady, compressible flow problems there are well defined curves or surfaces called characteristics along which physical disturbances such as shock waves propagate. For two or three spatial variables and time, waves propagate along characteristic surfaces, while for one spatial variable and time, waves propagate along characteristic curves. In the mathematics literature, characteristic curves and surfaces are called Cauchy characteristics. 

For two-dimensional supersonic flows containing shock waves, the characteristics are also called Mach lines. In compressible flows with waves, the equations defined throughout the flow simplify to equations specified on the characteristic curves (or surfaces) which reduces the number of spatial variables defining the problem. Restricting the governing PDEs to characteristics produces a new set of equations called compatibility equations. Equations (1), (2), and (3) describing the one-dimensional, unsteady, compressible flow of an isentropic gas, are a system of first-order PDEs in one space variable and time. Therefore, the blast wave problem may be reduced to a system of equations on the Mach lines governing the propagation of a shock wave.

To find the Mach lines and restrict the equations of motion to these lines, a linear combination of Eqs.\,(1), (2), and (3) is formed and recast in terms of total derivatives of the density, pressure and velocity:
\begin{equation*}
d\rho = \rho_t dt + \rho_x dx,
\end{equation*}
\begin{equation*}
dp = p_t dt + p_x dx,
\end{equation*}
and
\begin{equation*}
du = u_t dt + u_x dx.
\end{equation*}
The characteristic curves, denoted by \( dt/dx = \lambda \), are the coefficients of the time derivatives for the differentials of density, pressure, and velocity. A general method to derive the characteristic curves (or surfaces) as well as the restriction of the equations of motion to the characteristics is developed in Rusanov [27]. The restriction of the PDEs of motion to Mach lines for the blast wave problem reduces the problem to a set of ODEs in terms of the flow variables, \( \rho \), \( p \), and \( u \). Since the equations are cast in terms of the total derivatives of the flow variables, these equations are naturally interpreted as differential one-forms defined on submanifolds (the Mach lines) associated with the exterior differential system induced by the one-dimensional Euler equations. 

The abstract framework for exterior differential systems gives insight into the relationship between characteristics and symmetries. To see this relationship, let \( I \subset \Lambda^{\ast}(\Sigma)\) be a differential ideal representing the Euler Eqs.\,(1), (2), and (3), where \( \Sigma \) is a smooth surface defined as a subset of the jet space \( J^1({\mathbb{R}}^2, {\mathbb{R}^3}) \). A vector field \( V \in \Gamma (T\Sigma) \) (that is, \( V \) is section of the tangent bundle \( T\Sigma \) of \( \Sigma \)) is a Cauchy characteristic vector field for \( I \) if \( V \intprod \Psi \in I \) for all \( \Psi \in I \). Here \( V \intprod \Psi \) is the interior product of the vector field \( V \) with the differential form \( \Psi \). The key result is that if \( V \) is a Cauchy characteristic vector field of \( I \), it is also an infinitesimal symmetry of \( I \). Hence, the vector fields associated with the Mach lines are Lie symmetries of the Euler equations associated with the propagation of waves.

\subsection{Characteristics and Compatibility Equations}

The characteristic equations for one-dimensional, unsteady, compressible flow are given by
\begin{equation}
\frac{dx}{dt}= \frac{1}{\lambda} = u \ \ \ \text{ (Pathlines)},
\end{equation}
and
\begin{equation}
\frac{dx}{dt}= \frac{1}{\lambda_{\pm}} = u \pm a \ \ \ \text{ (Mach lines)},
\end{equation}
where \( a \) is the local speed of sound.
The associated compatibility equations are
\begin{equation}
dp - a^2 d\rho =0 \ \ \ \text{along Pathlines},
\end{equation}
and
\begin{equation}
dp \pm \rho a du + \alpha \frac{\rho u a^2}{x}dt = 0 \ \ \ \text{ along Mach lines},
\end{equation}
Zucrow and Hoffman, [28]. In Eq.\,(44), \( \alpha \) represents the coordinate system: \( \alpha = 0 \) cartesian coordinates, \( \alpha = 1\) cylindrical coordinates, and \( \alpha = 2\) spherical coordinates.

\subsection{Spherically Symmetric Potential Flow}

For a one-dimensional, spherically symmetric flow field, the velocity field will be irrotational at each point in the fluid. Hence, a potential function, \( \varphi \), exists such that \( u = \varphi_x \). For an unsteady, isentropic, compressible flow, the conservation equations, Eqs.\,(1) to (3), may be used to derive a second-order, hyperbolic, partial differential equation in terms of the potential function:
\begin{equation}
(a^2 - u^2)\varphi_{xx} - 2u\, \varphi_{xt} - \varphi_{tt} + 2\frac{a^2u}{x} = 0,
\end{equation} 
subject to the compressible Bernoulli equation,
\begin{equation}
\varphi_t + \frac{1}{2}u^2 + \int^{p}_{p_0} \frac{dp}{\rho} = 0.
\end{equation} 
Then restricting the potential equation (45) to the Mach lines (the characteristic curves) given by Eqs.\,(42), the compatibility equations for spherically symmetric, isentropic, compressible potential flow become:
\begin{equation}
(a^2 - u^2)du -(u \pm a)d\varphi_{t} + 2\frac{a^2u}{x}(u \pm a)dt = 0,
\end{equation}
Sauer, [29].

The compatibility equations (47) for spherically symmetric potential flow are equivalent to the compatibility equations associated with the Euler equations, Eqs.\ (44). To see the equivalence of the two sets of compatibility equations, compute the exterior derivative of the compressible Bernoulli equation to produce 
\begin{equation}
d\varphi_t = - u\, du - \frac{dp}{\rho}.
\end{equation} 
Combining Eq.\,(48) with Eqs.\,(47) then yields
\begin{equation}
dp \pm \rho a du + 2\frac{\rho u a^2}{x}dt = 0,
\end{equation}
along the Mach lines of Eqs.\,(44) as claimed.

\section{The Taylor Point Blast Wave as \\ Exterior Differential Systems}

\subsection{Similarity Solution of the Blast Wave}

Combining the functional forms of Eqs.\,(28) to (31) with expressions for \( R \) and \( dR/dt \) as well as the equations of conservation of mass, momentum and energy, it may shown that the following system of ODEs results modeling the blast wave:
\begin{equation}
\frac{d\phi}{d\eta}= \frac{1}{\eta - \phi}\Big( \frac{1}{\gamma}\frac{f'}{\psi} - \frac{3}{2} \phi\Big),
\end{equation}
\begin{equation}
\frac{df}{d\eta}=\frac{f(-3\eta +\phi (3 + \gamma / 2) -2\gamma \phi^2 / \eta)}{((\eta-\phi)^2-f/\psi)},
\end{equation}
and
\begin{equation}
\frac{d\psi}{d\eta}=\psi \frac{\phi' + 2\phi / \eta}{\eta - \psi},
\end{equation}
Taylor [1]; also see Sachdev [30]. Here the prime notation represents differentiation with respect to \( \eta \). Recall that Fig.\,3 shows plots of the numerical solutions of Eqs.\,(50), (51), and (52) for the limiting values of the strong shock conditions of Eqs.\,(4) to (6). 

The ODEs derived for the blast wave may then be cast as differential one-forms by defining:
\begin{equation}
\theta^ 1 := d\phi - \frac{1}{\eta - \phi}\Big( \frac{1}{\gamma}\frac{f'}{\psi} - \frac{3}{2} \phi\Big) d\eta,
\end{equation}
\begin{equation}
\theta^2 := df - \frac{f(-3\eta +\phi (3 + \gamma / 2) -2\gamma \phi^2 / \eta)}{((\eta-\phi)^2-f/\psi)} d\eta,
\end{equation}
and
\begin{equation}
\theta^3 := d\psi - \psi \frac{\phi' + 2\phi / \eta}{\eta - \psi} d\eta.
\end{equation}
Equations (53), (54), and (55) are applied to identify integral curves on the manifold \( \Sigma \) modeling solutions of the Euler equations (1), (2), and (3) with the coordinate cover \( (x,t,\rho, u, p) \). The following theorem (see Ivey and Landsberg [21]) gives conditions for the existence of solution curves to the ODEs above restricted to a submanifold of \( \Sigma \):  

\bigskip 

\noindent{{\bf{Theorem}} \emph{Let \( \Sigma \) be a \( C^{\infty} \) manifold of dimension \( m\), and let \( \theta^1, . . ., \theta^{m-1} \in \Lambda^1(\Sigma)\) be pointwise linearly independent in some neighborhood \( U \subset \Sigma \). Then through \( z \in U \) there exists a curve \( c : {\mathbb{R}} \longrightarrow U\), unique up to reparametrization, such that \( c^{\ast} (\theta^a) = 0 \) for \( 1 \le a \le m-1 \).}}

\bigskip

\noindent{The theorem guarantees that the system of ODEs modeling the self-similar motion of the blast wave will have a local solution represented by a curve \( c = (\phi (\eta), f(\eta), \psi(\eta))\) near a specified point \( z \in (\eta, \phi, f, \psi) \). The differential forms are zero along solution curves \( c \) of the ODEs used to define the forms.  Notice that the similarity transformation restricts the underlying surface \( (x, t, \rho, u, p)\) modeling the blast wave to a submanifold described by \( (\eta, \phi, f, \psi )\) with \( m = 4 \) in terms of \(m - 1 = 3 \) linearly independent one-forms, Eqs.\,(53), (54), and (55). 
 
\subsection{Exterior Forms Defined on the Blast Wave \\ Characteristics}

In Section 4, the Method of Characteristics was used to reduce the spherically symmetric Euler equations defined in space and time \( (x,t)\) to a system of ODEs (which may also be interpreted as an EDS) defined on the pathlines and Mach lines in the flow field. Restriction of the PDEs to the Mach lines projects the manifold \( \Sigma \) defined by the coordinates \( (x, t, \rho, u, p) \) onto the manifold defined by \( (t, \rho, u, p) \), and reduces the coordinate cover by one variable. The compatibility equations and the coordinates characteristic curves may be solved at the points were the characteristic curves intersect in space and time. 

The solution of the blast wave problem requires three linearly independent differential one-forms for the flow field properties \( (\rho, u, p) \). Along pathlines and Mach lines in the flow field, the governing one-forms are given by the compatibility equations (43) and (44):
\begin{equation}
\theta^1 = dp - a^2 d\rho, 
\end{equation}
\begin{equation}
\theta^2 = dp + \rho a du + \frac{2\rho u a^2}{x}dt,
\end{equation}
\begin{equation}
\theta^3 = dp - \rho a du + \frac{2\rho u a^2}{x}dt.
\end{equation}
The analysis of the EDS defined by Eqs.\,(56) to (58) does not use Eq.\,(41) for the pathline explicitly, because the pathline equation may be represented as a linear combination of the Mach lines of Eqs.\,(42). If this system of exterior differential forms is thought of a system of equations in \( (t, \rho, u, p) \), then the theorem of Section 5.1 guarantees that a local solution exists to the system of ODEs associated with Eqs.\,(56), (57), and (58) at points where the equations are not singular and where the characteristic curves intersect.

On the other hand, if the EDS defined by Eqs.\,(56), (57), and (58) is thought of as a system of differential one-forms defined on \( (x, t, \rho, u, p) \) and the equations for the Mach lines are not used initially to restrict the number of independent variables to one, the theorem of Section 5.1 does not apply. For this case, the problem coordinates are \( (x, t, \rho, u, p) \) and an integration theorem must be applied that holds for more than one independent variable.  

The Frobenius theorem of differential geometry gives integrability conditions for a system of differential one-forms on a surface. This theorem extends the concept of integrating factors from a single equation to a system of equations. The Frobenius theorem has at least two natural forms: one which gives a functional representation of differential one-forms, and one which expresses an equivalent concept in terms of the closure properties of a differential ideal defined by differential one-forms. Following Ivey and Landsberg [21], the Frobenius theorem in terms of differential one-forms is:

\bigskip

\noindent{{\bf{Theorem}} \emph{Let \( \Sigma \) be a \( C^{\infty} \) manifold of dimension \( m\), and let \( \theta^1, . . ., \theta^{m-n} \in \Lambda^1(\Sigma)\) be pointwise linearly independent. If there exist one-forms \( \alpha^i_j \in \Lambda^1(\Sigma) \) such that \( d\theta^j = \alpha^j_i \wedge \theta^i \) for all \( j \), then through each point \( z \in \Sigma \) there exists a maximal connected \( n \)-dimensional manifold \( i : N \hookrightarrow \Sigma \) such that \( d\theta^j = \alpha^j_i \wedge \theta^i \) for \( 1 \le j \le m-n \). This manifold is unique, in the sense that any other such connected submanifold through \( z \) is a subset of \( i(N) \).}}

\bigskip

The Frobenius theorem will be applied to show that the EDS defined by Eqs.\,(56) to (58) is integrable. Notice that the conditions
\begin{equation}
d\theta^j = \alpha^j_i \wedge \theta^i \ \ \ \text{ for } \ \ \ i, j =1,..., m-n,
\end{equation}
are equivalent to 
\begin{equation}
\Omega \wedge d\theta^j = 0 \ \ \text{for} \ \ j =1,..., m-n,
\end{equation}
where
\begin{equation}
\Omega = \theta^1 \wedge \theta^2 \wedge \theta^3 \ne 0.
\end{equation}
Moreover, it is shown in Flanders [31] that Eqs.\,(59) assure the existence functions \( f^i_j \) and \( g^j \) for \( i,j = 1, ...,m-n \) such that
\begin{equation}
\theta^i = f^i_1 \, dg^1 + \cdots + f^i_{m-n} \, dg^{m-n}.
\end{equation}
Equations (62) imply that integrating factors exist which allow a EDS to be represented in integral form.

The Frobenius theorem is satisfied for the EDS modeling the blast wave if the one-forms of Eqs.\,(56) to (58) are linearly independent
\begin{equation}
\theta^1 \wedge \theta^2 \wedge \theta^3 \ne 0,
\end{equation}
and if
\begin{equation*}
\theta^1 \wedge \theta^2 \wedge \theta^3 \wedge d\theta^i = 0, \ \ \text{for} \ \ i =1, 2, 3.
\end{equation*}
are satisfied using Eqs\,(56) to (58). Direct calculation of the wedge product produces:
\begin{multline}
\theta^1 \wedge \theta^2 \wedge \theta^3 = -4\frac{\rho^2 u a^3}{x}\, dt \wedge du \wedge dp -2\rho a^3 \, d\rho \wedge du \wedge dp \\
-4\frac{\rho^2 u a^5}{x}\, dt \wedge d\rho \wedge du \ne 0.
\end{multline}
Moreover, calculation of the exterior derivative of \( \theta^1, \theta^2 \), and \( \theta^3 \) gives:
\begin{equation}
d\theta^1 = -2a\,da \wedge d\rho,
\end{equation}
\begin{multline}
d\theta^2 = a\,d\rho \wedge du + \rho \, da \wedge du - 2\frac{ua^2}{x} dt \wedge d\rho - 2\frac{\rho a^2}{x} dt \wedge du \\
- 4\frac{\rho u a}{x} dt \wedge da - 2\frac{\rho u a^2}{x^2} dx \wedge dt,
\end{multline}
and
\begin{multline}
d\theta^3 = a\,d\rho \wedge du - \rho \, da \wedge du - 2\frac{ua^2}{x} dt \wedge d\rho - 2\frac{\rho a^2}{x} dt \wedge du \\
- 4\frac{\rho u a}{x} dt \wedge da - 2\frac{\rho u a^2}{x^2} dx \wedge dt.
\end{multline}
To simplify Eqs.\,(65) to (67), recall that for the speed of sound \( a^2 = \gamma p/\rho \), so that 
\begin{equation}
2a\,da = \frac{\gamma}{\rho}dp - \frac{\gamma p}{\rho^2 }d\rho.
\end{equation}
Then combining Eq.\,(68) with Eqs.\,(65) to (67) yields:
\begin{equation}
d\theta^1 = \frac{\gamma}{\rho} d\rho \wedge dp,
\end{equation}
\begin{multline}
d\theta^2 = a \, d\rho \wedge du - \frac{\gamma}{2a} du \wedge dp 
- \frac{\gamma p}{2 a \rho} d\rho \wedge du \\
- 2\frac{ua^2}{x} dt \wedge d\rho  
- 2\frac{\rho a^2}{x} dt \wedge du 
- 2\frac{\gamma u}{x} dt \wedge dp \\
+ 2\frac{\gamma u p}{\rho x} dt \wedge d\rho
- 2\frac{\rho u a^2}{x^2} dx \wedge dt,
\end{multline}
\begin{multline}
d\theta^3 = a \, d\rho \wedge du + \frac{\gamma}{2a} du \wedge dp + \frac{\gamma p}{2 a \rho} d\rho \wedge du \\
- 2\frac{ua^2}{x} dt \wedge d\rho  - 2\frac{\rho a^2}{x} dt \wedge du
- 2\frac{\gamma u}{x} dt \wedge dp \\
+ 2\frac{\gamma u p}{\rho x} dt \wedge d\rho
- 2\frac{\rho u a^2}{x^2} dx \wedge dt.
\end{multline}
Combining Eq.\,(64) with the wedge product of Eqs.\,(69), (70), and (71) implies
\begin{equation}
\theta^1 \wedge \theta^2 \wedge \theta^3 \wedge d\theta^1= 0,
\end{equation}
and
\begin{multline}
\theta^1 \wedge \theta^2 \wedge \theta^3 \wedge d\theta^i = \\
4\frac{\rho^2 u a^5}{x^2} dx \wedge dt \wedge d\rho \wedge du \wedge dp, \ \ \text{for} \ \ i =2,3.
\end{multline}
Finally, restricting the above differential form to the Mach lines, \( dx = (u \pm a) dt \), implies that the right-hand-side of Eqs.\,(73) are linearly dependent, so it follows that:
\begin{multline}
\theta^1 \wedge \theta^2 \wedge \theta^3 \wedge d\theta^i = \\
4\frac{\rho^2 u a^5(u \pm a)}{x^2} dt \wedge dt \wedge d\rho \wedge du \wedge dp = 0 \\ \text{for} \ \ i =2,3.
\end{multline}
Hence, the conditions of the Frobenius theorem are satisfied for the exterior differential system defined by Eqs.\,(56) to (58) along the Mach lines and the compatibility equations modeling the blast wave are integrable at the nonsingular points where the characteristic curves intersect. 

\section{Summary and Discussion}

This article presented a brief historical overview of G.\,I.\,Taylor's solution of the point blast wave problem and its applications to the analysis of the Trinity atomic test in 1945. The blast wave solution formulated by Taylor, which used both dimensional analysis and similarity techniques to simplify the equations of motion, resulted in a powerful mathematical model that yielded fundamental insight into the physics of blast waves. This article further showed that the solution methods applied by Taylor are examples of the practical manifestation of the symmetry properties associated with the physics of shock wave propagation. 

The key points of this study may be summarized as:

\noindent{1. Lie group methods were applied to derive Taylor's famous two-fifths dimensional result relating the position of the blast wave \( R \) to the time \( t \) elapsed in the explosion as well as the total energy released. The use of Lie group methods demonstrates that a basic relationship exists between the geometry (or symmetry) and physics of wave propagation through the equations of motion.} 

\noindent{2. The set of self-similar ordinary differential equations derived by Taylor to model the point blast wave were shown to be equivalent to an exterior differential system that has a local solution for the velocity, pressure, and density along a curve representing the motion of the blast wave.}

\noindent{3. The method of characteristics for one-dimensional, unsteady, compressible flow was shown to be equivalent to the method of characteristics for one-dimensional, unsteady, compressible, potential flow. The equations of the method of characteristics for spherically symmetric, unsteady, compressible flow were then shown to yield a natural exterior differential system modeling the point blast wave along Mach lines.}

\noindent{4. The exterior differential system modeling the point blast wave derived from the method of characteristics was shown to be locally integrable, that is, to have local solutions for velocity, pressure, and density at the points in space and time where the characteristic curves intersect.} 

This article has shown that the solution of the point blast wave problem is directly related to the symmetry properties of the motion of the underlining shock wave. The Lie group symmetry techniques generalize to the study of exterior differential systems which provide a theoretical framework for analyzing shock wave propagation problems in multiple spatial dimensions with the method of characteristics. Historically, the method of characteristics has been used mainly to compute compressible flow problems in one space dimension and time because the characteristics reduce to planar curves and the equations of motion reduce to ODEs.

The geometry (or symmetry) and physics of complex shock wave phenomena may be studied in detail by applying the methods of EDS to guide and interpret numerical simulations that have higher geometric precision and greater physical fidelity than analytic models. For example, an inviscid flow field with low levels of vorticity may be approximated as a potential flow, which reduces a system of first-order PDEs for the velocity vector to a single second-order PDE for a velocity potential. Recall that the point blast wave may be modeled with the potential equation (45), which is equivalent to the Euler equations (1), (2), and (3) for spherically symmetric, isentropic, compressible flow. Bryant, et al.\,[32] have used the theory of exterior differential systems to analyze rigorously the integrability of a general class of second-order PDEs which includes the equations for compressible potential flow. A detailed understanding of compressible potential flow would provide insight into the basic physics of shock wave motion and help explain and quantify more complex physical effects found in numerical simulations such as the interaction of shock waves and the generation of vorticity. 

\section*{Acknowledgements}

The authors would like to thank Richard Moore of the Atomic Weapons Establishment, as well as Ralph Menikoff, Christopher Triola, Joe Schmidt, Cory Ahrens, Len Margolin, and Mark Chadwick of the Los Alamos National Laboratory for their many helpful suggestions in the development of this article. 
This work was supported by the US Department of Energy through
the Los Alamos National Laboratory. Los Alamos National Laboratory is
operated by Triad National Security, LLC, for the National Nuclear Security
Administration of the US Department of Energy under Contract No.
89233218CNA000001.

\end{document}